\documentclass[prb,amsmath,amsfonts,amssymb,unsortedaddress,floatfix]{revtex4-1}
\usepackage{dcolumn}
\usepackage{graphicx}
\usepackage{bm}
\usepackage{color}

\begin{document}
\title{Pair extended coupled cluster doubles}
\author{Thomas M. Henderson}
\affiliation{Department of Chemistry, Rice University, Houston, TX 77005-1892}
\affiliation{Department of Physics and Astronomy, Rice University, Houston, TX 77005-1892}

\author{Ireneusz W. Bulik}
\affiliation{Department of Chemistry, Rice University, Houston, TX 77005-1892}

\author{Gustavo E. Scuseria}
\affiliation{Department of Chemistry, Rice University, Houston, TX 77005-1892}
\affiliation{Department of Physics and Astronomy, Rice University, Houston, TX 77005-1892}
\date{\today}

\begin{abstract}
The accurate and efficient description of strongly correlated systems remains an important challenge for computational methods.  Doubly occupied configuration interaction (DOCI), in which all electrons are paired and no correlations which break these pairs are permitted, can in many cases provide an accurate account of strong correlations, albeit at combinatorial computational cost.  Recently, there has been significant interest in a method we refer to as pair coupled cluster doubles (pCCD), a variant of coupled cluster doubles in which the electrons are paired.  This is simply because pCCD provides energies nearly identical to those of DOCI, but at mean-field computational cost (disregarding the cost of the two-electron integral transformation).  Here, we introduce the more complete pair extended coupled cluster doubles (pECCD) approach which, like pCCD, has mean-field cost and reproduces DOCI energetically.  We show that unlike pCCD, pECCD also reproduces the DOCI wave function with high accuracy.  Moreoever, pECCD yields sensible albeit inexact results even for attractive interactions where pCCD breaks down.
\end{abstract}
\maketitle

\section{Introduction}
While the desciption of the ground state of weakly correlated systems is by now fairly routine, the same cannot be said for strongly correlated problems.  Because coupled cluster theory\cite{Paldus1999,Bartlett2007,BartlettShavitt} offers exceptional accuracy for the description of weak dynamic correlations, we would like to use some variant of coupled cluster theory as well for the strong static correlations; in this way one could seamlessly merge the two ideas to provide a more powerful technique which should be accurate for both regimes.  Unfortunately, the construction of coupled cluster techniques for strongly correlated systems is a work in progress.

A particularly interesting recent development is the notion of pair coupled cluster doubles (pCCD),\cite{Limacher2013,Limacher2014,Tecmer2014,Boguslawski2014,Stein2014,Henderson2014b} which takes the simple coupled cluster wave function and makes the dramatic simplification that the only allowed excitations are of a paired form in which two electrons are removed from the same spatial orbital and placed in some other spatial orbital.  In other words, pCCD is coupled cluster doubles restricted to include only seniority zero determinants, where the seniority of a determinant is the number of singly occupied spatial orbitals.  This restriction greatly decreases the cost of the coupled cluster calculation to mean field or $\mathcal{O}(M^3)$ if one ignores the two-electron integral transformation, but rather paradoxically, despite this simplification the pCCD wave function yields results very close to the doubly occupied configuration interaction (DOCI),\cite{Allen1962,Smith1965,Weinhold1967,Veillard1967,Couty1997,Kollmar2003,Bytautas2011} which includes such pair excitations to all excitation levels.  This DOCI method is not new, and includes many powerful geminal wave functions, including the antisymmetrized geminal power (APG), the antisymmetric product of strongly orthogonal geminals (APSG), and many others.  For many problems of interest, DOCI is able to describe the basics of the strong correlations, and to the extent that pCCD reproduces DOCI, so too does pCCD.

We note, however, that the coincidence between pCCD and DOCI is not entirely universal.  For the attractive pairing Hamiltonian
\begin{equation}
H = \sum_p \epsilon_p \, \left(a_{p_\uparrow}^\dagger \, a_{p_\uparrow} + a_{p_\downarrow}^\dagger \, a_{p_\downarrow}\right)
  - G \sum_{pq} a_{p_\uparrow}^\dagger \, a_{p_\downarrow}^\dagger \, a_{q_\downarrow} \, a_{q_\uparrow}
\label{Eqn:PairingHam}
\end{equation}
we observe that as the pairing strength $G$ becomes larger and the mean-field solution develops an instability toward a number symmetry broken Hartree-Fock-Bogoliubov state, pCCD breaks down dramatically, overcorrelating wildly before eventually returning complex energies.\cite{Henderson2014}  This suggests that pCCD may not be able to describe the kinds of strong correlations needed to model superconductivity, for example.  

Moreover, even when the pCCD energy is accurate, the wave functions may be less so.  In coupled cluster theory, the Hamiltonian is similarity transformed, yielding a non-Hermitian effective Hamiltonian $\bar{H} = \exp(-T) \, H \, \exp(T)$.  Because $\bar{H}$ is non-Hermitian, it has different left- and right-hand eigenvectors:
\begin{subequations}
\begin{align}
\bar{H} |0\rangle &= E | 0\rangle,
\\
\langle 0| (1+Z) \, \bar{H} &= E \langle 0 | (1+Z)
\end{align}
\end{subequations}
where $Z$ creates excitations to the left and $|0\rangle$ is the mean-field reference.  This structure in turn translates to different left- and right-hand wave functions for the original Hamiltonian, which in pCCD are taken to be
\begin{subequations}
\begin{align}
\langle \mathcal{L}_\mathrm{pCCD} | &= \langle 0 | (1+Z) \, \mathrm{e}^{-T},
\\
| \mathcal{R}_\mathrm{pCCD} \rangle &= \mathrm{e}^T | 0\rangle,
\end{align}
\end{subequations}
where $T$ and $Z$ respectively create pair excitations and pair de-excitations when acting to the right.  In Ref. \onlinecite{Henderson2014b} we and others showed that the overlap
\begin{equation}
S = \langle \mathcal{L}_\mathrm{pCCD} | \mathrm{DOCI} \rangle \langle \mathrm{DOCI} | \mathcal{R}_\mathrm{pCCD} \rangle
\end{equation}
is close to unity, but this does not separately test the pCCD left-hand and right-hand wave functions and, as we will see below, while the pCCD right-hand wave function is close to the DOCI state, the same is not necessarily true of the pCCD left-hand wave function.  This can cause the pCCD density matrices to be less accurate approximations of the actual DOCI density matrices than we would like.

This manuscript introduces the pair extended coupled cluster method (pECCD) which seeks to remedy these deficiencies.  Where pCCD is the seniority zero version of coupled cluster doubles, pECCD is the seniority zero version of the extended coupled cluster doubles method of Arponen and Bishop.\cite{Arponen1983,Arponen1987,Arponen1987b,Piecuch1999}  While extended coupled cluster has not seen a great deal of use due to its large computational cost ($\mathcal{O}(M^{10})$ for extended coupled cluster doubles\cite{Fan2006}), pECCD has the same mild scaling with system size displayed by pCCD, though with a rather larger prefactor.  We will sketch the pCCD method in Sec. \ref{Sec:Theory} and provide a few results in Sec. \ref{Sec:Results} to show the basic performance of the method before providing our conclusions and prospects for further development in Sec. \ref{Sec:Conclusions}.  We should note that the equations needed for efficient computation of the pECCD energy and wave function amplitudes are exceedingly lengthy despite their low computational cost; here, we provide a simpler but computationally more demanding expression for the energy (from which the amplitude equations follow) and have not included the amplitude equations at all.  Accordingly, the code needed to solve pECCD more efficiently will be made available upon request.

\section{Theory
\label{Sec:Theory}}
The basics of the pair extended coupled cluster method are simple.  We define a pair excitation operator
\begin{equation}
T = \sum_{ai} t_{ia} \, P_a^\dagger \, P_i
\label{DefT}
\end{equation}
and a pair de-excitation operator
\begin{equation}
Z = \sum_{ai} z_{ai}\, P_i^\dagger \, P_a
\label{DefZ}
\end{equation}
where the pair creation operator $P_p^\dagger$ is given by
\begin{equation}
P_p^\dagger = a_{p_\uparrow}^\dagger \, a_{p_\downarrow}^\dagger
\label{DefP}
\end{equation}
and the pair annihilation operator $P_p$ is its adjoint.  Note that these operators are nilpotent ($P_p^2 = 0$) and that more general pairing schemes than this simple singlet pairing within the same spatial orbital are possible.

Regardless, having defined an excitation operator $T$ and a de-excitation operator $Z$, the pECCD energy is given by
\begin{subequations}
\begin{align}
E_\mathrm{pECCD} 
 &= \langle 0| \mathrm{e}^Z \, \mathrm{e}^{-T} \, H \, \mathrm{e}^T \, \mathrm{e}^{-Z} | 0\rangle
\\
 &= \langle 0| (1 + Z + \frac{1}{2} \, Z^2 + \frac{1}{6} \, Z^3) \, \mathrm{e}^{-T} \, H \, \mathrm{e}^T | 0\rangle
\end{align}
\end{subequations}
where in the second line we have used the facts that the de-excitation operator $Z$ annihilates the vacuum to the right and that the similarity transformed Hamiltonian $\bar{H} = \mathrm{e}^{-T} \, H \, \mathrm{e}^T$ is a six-body operator that creates up to hextuple excitations to the right, so that no more than hextuple de-excitations (created by $Z^3$) are needed for the case under consideration.  The amplitudes $t_{ia}$ and $z_{ai}$ are obtained by solving
\begin{equation}
0 = \frac{\partial E_\mathrm{pECCD}}{\partial t_{ia}} = \frac{\partial E_\mathrm{pECCD}}{\partial z_{ai}}.
\end{equation}

We will take the physical Hamiltonian to be
\begin{equation}
H = \sum_{pq} \sum_\sigma \langle p | h | q \rangle \, a_{p_\sigma}^\dagger \, a_{q_\sigma}
  + \frac{1}{2} \, \sum_{pqrs} \, \sum_{\sigma\eta} \langle pq| v | rs \rangle \, a_{p_\sigma}^\dagger \, a_{q_\eta}^\dagger \, a_{s_\eta} \, a_{r_\sigma}
\end{equation}
where $\langle p | h | q \rangle$ and $\langle pq | v | rs \rangle$ are one- and two-electron integrals; note that the two-electron integrals here are not antisymmetrized.  Greek letters index spin while Latin letters index spatial orbitals, with $i, j, k \ldots$ denoting occupied orbitals, $a, b, c, \ldots$ denoting virtual orbitals, and $p, q, r, \ldots$ denoting arbitrary orbitals.

We can greatly simplify the derivation of the pECCD energy (and therefore amplitude equations) by replacing the physical Hamiltonian $H$ with the portion $H^{\delta\Omega=0}$ which preserves seniority.  For a two-body Hamiltonian, we would have
\begin{equation}
H = H^{\delta\Omega=0} + H^{\delta\Omega=2} + H^{\delta\Omega=4}
\end{equation}
where $H^{\delta\Omega=2}$ and $H^{\delta\Omega=4}$ couple determinants whose seniorities differ by two and by four, respectively.  Determinants differing by an odd seniority have different electron numbers, while determinants differing by an even seniority greater than four differ by a triple excitation or higher (and thus cannot be coupled by a two-body operator).  Because every spatial orbital in pCCD or pECCD has seniority zero (\textit{i.e.} all orbitals are either doubly occupied or empty), it will suffice for our purposes to determine what we will call $H_0^{\delta\Omega=0}$, which is the part of $H^{\delta\Omega=0}$ that preserves the seniority of each orbital.  We caution, however, that $H_0^{\delta\Omega=0}$ omits terms which change the seniority of individual orbitals while preserving the total seniority; these terms do not contribute to closed-shell pCCD or pECCD but could contribute to a kind of ROHF-based generalization of pCCD or pECCD.

We derive our expression for $H_0^{\delta\Omega=0}$ in the appendix and merely quote it here:
\begin{equation}
H_0^{\delta\Omega=0}
 = \sum_p h_p \, N_p
 + \frac{1}{4} \, \sum_{p \neq q} w_{pq} \, N_p \, N_q
 + \sum_{pq} v_{pq} \, P_p^\dagger \, P_q
 + \sum_{p \neq q} K_{pq} \, \vec{S}_p \cdot \vec{S}_q
\label{DefH0}
\end{equation}
where the necessary integrals are
\begin{subequations}
\begin{align}
h_p &= \langle p | h | p\rangle,
\\
v_{pq} &= \langle pp | v | qq \rangle,
\\
w_{pq} &= 2 \, \langle pq | v | pq \rangle - \langle pq | v | qp \rangle,
\\
K_{pq} &= -\langle pq | v | qp \rangle
\end{align}
\end{subequations}
and where the number and spin operators are given by
\begin{subequations}
\begin{align}
N_p &= \sum_\sigma a_{p_\sigma}^\dagger \, a_{p_\sigma},
\label{DefN}
\\
\vec{S}_p &= \frac{1}{2} \, \sum_{\xi\eta} a_{p_\xi}^\dagger \, \left(\vec{\bm{\sigma}}\right)_{\xi\eta} \, a_{p_\eta}
\label{DefSpin}
\end{align}
\end{subequations}
where $\vec{\bm{\sigma}}$ is the vector of Pauli matrices.  The Heisenberg-like term $\sum_{p \neq q} K_{pq} \, \vec{S}_p \cdot \vec{S}_q$ will not contribute in our closed-shell case because closed-shell orbitals have spin zero.  The number operator and pair creation and annihilation operators satisfy SU(2) commutation relationships
\begin{subequations}
\begin{align}
[P_p, P_q^\dagger] &= \delta_{pq} \, \left(1 - N_p\right),
\\
[N_p, P_q] &= -2 \, \delta_{pq} \, P_q.
\end{align}
\label{DefSU2}
\end{subequations}
One can verify that
\begin{equation}
\langle 0| H_0^{\delta\Omega=0} | 0\rangle = \langle 0| H |0\rangle
\end{equation}
where $|0\rangle$ is an RHF or ROHF determinant (or, more precisely, any determinant which uses the same spatial orbitals for different spins).

Given the seniority-preserving Hamiltonian, the pECCD energy can be obtained simply by computing the pECCD density matrices.  As discussed in Ref. \onlinecite{Henderson2014b}, the density matrices of seniority zero methods are sparse, and in fact the only non-zero elements are precisely those we need to evaluate the expectation value of $H_0^{\delta\Omega=0}$.  We may of course also use density matrices to take other expectation values; in particular, they can be used in orbital optimization.\cite{Henderson2014b}  In this work, we will use orbitals optimized for pCCD, because in our experience the pCCD- and pECCD-optimized orbitals are virtually identical and the pCCD orbital optimization is somewhat less expensive because the pCCD density matrices are simpler to compute.

To compute density matrix elements, we define a pECCD expectation value via
\begin{equation}
\langle \mathcal{O} \rangle = \langle 0| \mathrm{e}^{Z} \, \mathrm{e}^{-T} \, \mathcal{O} \, \mathrm{e}^T | 0 \rangle.
\end{equation}
Given this expectation value, we find that the density matrices are given by
\begin{subequations}
\begin{align}
\langle N_i \rangle &= 2 \, \left(1 - \sum_a t_{ia} \, z_{ai}\right),
\\
\langle N_a \rangle &= 2 \, \sum_i t_{ia} \, z_{ai},
\\
\langle N_i \, N_j \rangle
 &= 4 \left(1 
  - \sum_a \left(t_{ia} \, z_{ai} + t_{ja} \, z_{aj}\right)
  + \delta_{ij} \, \sum_a t_{ia} \, z_{ai}
  + \bar{\delta}_{ij} \, \sum_{a \neq b} t_{ia} \, t_{jb} \, \left(z_{ai} \, z_{bj} + z_{bi} \, z_{aj}\right)
  \right),
\\
\langle N_a \, N_b\rangle
 &= 4 \, \left(\delta_{ab} \, \sum_i z_{ai} \, t_{ia} 
  +       \bar{\delta}_{ab} \, \sum_{i \neq j} t_{ia} \, t_{jb} \, \left(z_{ai} \, z_{bj} + z_{bi} \, z_{aj}\right)\right),
\\
\langle N_j \, N_a\rangle
 &= 4 \, \left(\sum_{i \neq j} t_{ia} \, z_{ai}
  - \sum_{i \neq j} \, \sum_{b \neq a} t_{ia} \, t_{jb} \, \left(z_{ai} \, z_{bj} + z_{bi} \, z_{aj}\right)\right),
\\
\langle P_a^\dagger \, P_i\rangle
 &= z_{ai},
\\
\langle P_a^\dagger \, P_b\rangle
 &= \sum_i z_{ai} \, t_{ib} - 2 \, \bar{\delta}_{ab} \, \sum_{i \neq j} z_{ai} \, z_{bj} \, t_{ib} \, t_{jb},
\\
\langle P_i^\dagger \, P_j\rangle
 &= \sum_a t_{ia} \, z_{aj}
  + \delta_{ij} \, \left(1 - 2 \, \sum_a t_{ia} \, z_{ai}\right)
  - 2\,  \bar{\delta}_{ij} \, \sum_{a \neq b} t_{ia} \, t_{ib} \, z_{ai} \, z_{bj},
\\
\langle P_i^\dagger \, P_a\rangle
 &= t_{ia}
   - 2 \, \sum_{j \neq i} t_{ia} \, t_{ja} \, z_{aj}
  - 2 \, \sum_{b} t_{ib} \, t_{ia} \, z_{bi}
  + \sum_{bj} t_{ib} \, t_{ja} \, z_{bj}
  - 2 \, \sum_{b \neq a} \, \sum_{j \neq k} t_{ib} \, t_{ja} \, t_{ka} \, z_{aj} \, z_{bk}
\\
 &+ 4 \, \sum_{b \neq a} \sum_{j \neq i} t_{ia} \, t_{ib} \, t_{ja} \, \left(z_{ai} \, z_{bj} + z_{bi} \, z_{aj}\right)
  - 2 \, \sum_{b \neq c} \sum_{j \neq i} t_{ib} \, t_{ic} \, t_{ja} \, z_{bi} \, z_{cj}
\nonumber
\\
 &+ 2 \, \sum_{b \neq c \neq a} \sum_{j \neq k \neq i} t_{ib} \, t_{ic} \, t_{ja} \, t_{ka} \, \left(z_{ai} \, z_{bj} \, z_{ck} + 2 \, z_{aj} \, z_{bk} \, z_{ci}\right)
\nonumber
\end{align}
\label{DefDMats}
\end{subequations}
where $\bar{\delta}_{pq} = 1 - \delta_{pq}$.  Then the energy is just given by
\begin{align}
E_\mathrm{pECCD}
 &= \sum h_i \, \langle N_i \rangle
  + \sum h_a \, \langle N_a \rangle
  + \frac{1}{4} \, \sum_{i \neq j} w_{ij} \, \langle N_i \, N_j \rangle
  + \frac{1}{4} \, \sum_{a \neq b} w_{ab} \, \langle N_a \, N_b \rangle
  + \frac{1}{2} \, \sum_{ja} w_{ja} \, \langle N_j \, N_a \rangle
\\
 &+ \sum_{ij} v_{ij} \, \langle P_i^\dagger \, P_j \rangle
  + \sum_{ia} \left(v_{ia} \, \langle P_i^\dagger \, P_a \rangle + v_{ai} \, \langle P_a^\dagger \, P_i \rangle\right)
  + \sum_{ab} v_{ab} \, \langle P_a^\dagger \, P_b \rangle
\nonumber
\end{align}
where we have disregarded the Heisenberg Hamiltonian term as its expectation value vanishes.  One can verify that the pCCD energy is properly reproduced by taking the pECCD energy and omitting terms of $\mathcal{O}(Z^2)$ or $\mathcal{O}(Z^3)$.

While constructing the density matrices given in Eqn. \ref{DefDMats} would appear to require $\mathcal{O}(N^6)$ computational time, the summation restrictions can be lifted by adding and removing terms which, with sufficient exertion, allows one to evaluate the density matrices and thus the energy in $\mathcal{O}(N^3)$ time after introducing intermediates.  For example, we may write
\begin{subequations}
\begin{align}
\sum_{a \neq b} t_{ia} \, t_{ib} \, z_{ai} \, z_{bj}
 &= \sum_{ab} t_{ia} \, t_{ib} \, z_{ai} \, z_{bj}
  - \sum_a    t_{ia} \, t_{ia} \, z_{ai} \, z_{aj}
\\
 &= x_{ii} \, x_{ij} - Y_{ij}
\end{align}
\end{subequations}
in terms of intermediates 
\begin{subequations}
\begin{align}
x_{ij} &= \sum_a t_{ia} \, z_{aj},
\\
Y_{ij} &= \sum_a t_{ia} \, t_{ia} \, z_{ai} \, z_{aj}.
\end{align}
\end{subequations}
Similarly, the amplitude equations can be solved in $\mathcal{O}(N^3)$ operations with the appropriate definition of intermediates.  We have checked the correctness of our $\mathcal{O}(N^3)$ implementation by comparison to the explicit $\mathcal{O}(N^6)$ result for random input $T$ and $Z$ amplitudes and integrals $h$, $v$, and $w$, and have verified our $\mathcal{O}(N^3)$ amplitude equations by comparing analytic and numerical derivatives of $E_\mathrm{pECCD}$ for random inputs.

We should emphasize that while the pCCD and pECCD density matrices both adopt a quasi-diagonal form, the pECCD two-particle density matrices are much more complicated.  This is simply because the pECCD left-hand wave function (see below) is more sophisticated; it contains excitations to all even orders and thereby has more flexibility in fitting DOCI than does pCCD, even though pCCD and pECCD have the same number of parameters to optimize.

\section{Results
\label{Sec:Results}}
Following Ref. \onlinecite{Henderson2014b}, we will compare the pCCD, pECCD, and DOCI energies for a variety of systems, defining, for example,
\begin{equation}
\Delta E_\mathrm{pECCD} = E_\mathrm{pECCD} - E_\mathrm{DOCI},
\label{DefDE}
\end{equation}
and will also assess the quality of the pCCD and pECCD wave functions by evaluating
\begin{equation}
S = \langle \mathcal{L} | \mathrm{DOCI} \rangle \, \langle \mathrm{DOCI} | \mathcal{R} \rangle,
\label{DefS}
\end{equation}
where for both pCCD and pECCD the right-hand wave function is
\begin{equation}
|\mathcal{R}\rangle = \mathrm{e}^T | 0 \rangle
\end{equation}
while the left-hand wave functions are
\begin{subequations}
\begin{align}
\langle \mathcal{L}_\mathrm{pECCD} | &= \langle 0| \mathrm{e}^{Z} \, \mathrm{e}^{-T} 
\\
\langle \mathcal{L}_\mathrm{pCCD} | &= \langle 0| (1+Z) \, \mathrm{e}^{-T}
\end{align}
\end{subequations}
for pECCD and pCCD, respectively.  Recall that
\begin{equation}
1 = \langle \mathcal{L} | \mathcal{R} \rangle = \langle \mathcal{L} | \mathrm{DOCI} \rangle \, \langle \mathrm{DOCI} | \mathcal{R} \rangle + \sum_k \langle \mathcal{L} | \mathrm{DOCI}_k \rangle \, \langle \mathrm{DOCI}_k | \mathcal{R} \rangle
\end{equation}
where $|\mathrm{DOCI}_k\rangle$ is the $k^{th}$ excited DOCI state; thus, we will have $S \approx 1$ provided that the DOCI excited states have minimal overlap with either the left-hand or right-hand state of pECCD or pCCD.  It will also prove fruitful to look in more detail, however, at the individual left- and right-hand overlaps
\begin{subequations}
\begin{align}
S_L &= \mathcal{N}_L \, \langle \mathcal{L} | \mathrm{DOCI} \rangle,
\label{DefSL}
\\
S_R &= \mathcal{N}_R \, \langle \mathrm{DOCI} | \mathcal{R} \rangle,
\label{DefSR}
\end{align}
\label{DefSLR}
\end{subequations}
where the normalization constants $\mathcal{N}_L$ and $\mathcal{N}_R$ are such that the left- and right-hand states are individually normalized to unity so that, for example,
\begin{equation}
\mathcal{N}_R^2 \, \langle \mathcal{R} | \mathcal{R} \rangle = \mathcal{N}_R^2 \, \langle 0 | \mathrm{e}^{T^\dagger} \, \mathrm{e}^T | 0 \rangle = 1.
\end{equation}
These allow us to assess separately the quality of the left- and right-hand wave functions.

As Ref. \onlinecite{Henderson2014b} makes clear, with orbital optimization, pCCD is exact for two-electron singlets.  The same is of course true for pECCD.  Thus, both are exact for H$_2$ and nearly exact for LiH, with little to distinguish the two approaches in the latter case.  It will therefore be more fruitful to focus on systems with more strongly correlated electrons.

\begin{figure}[t]
\includegraphics[width=0.48\textwidth]{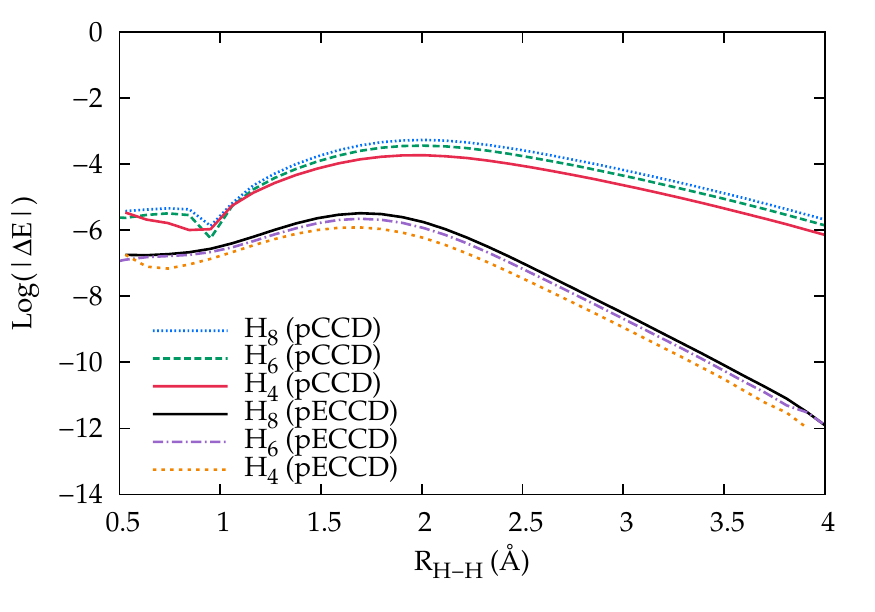}
\hfill
\includegraphics[width=0.48\textwidth]{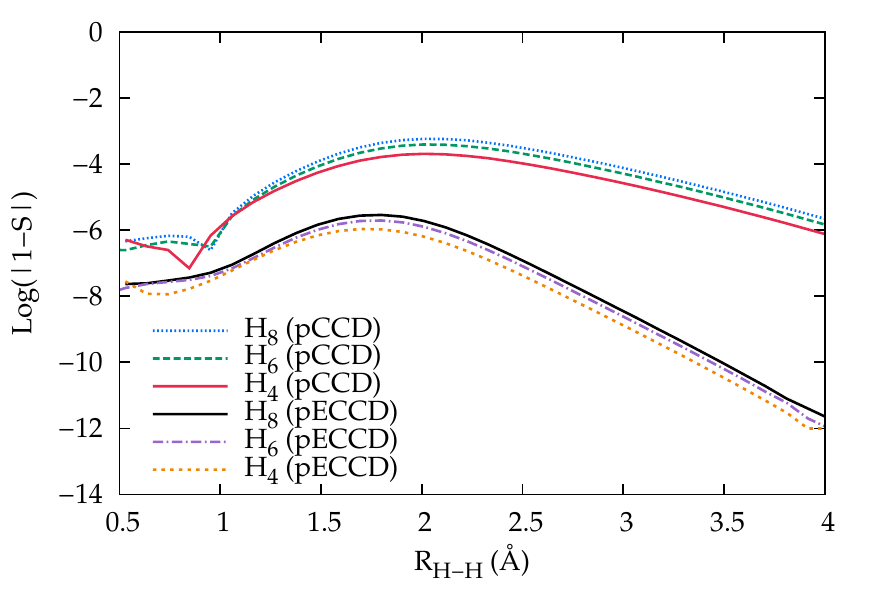}
\\
\includegraphics[width=0.48\textwidth]{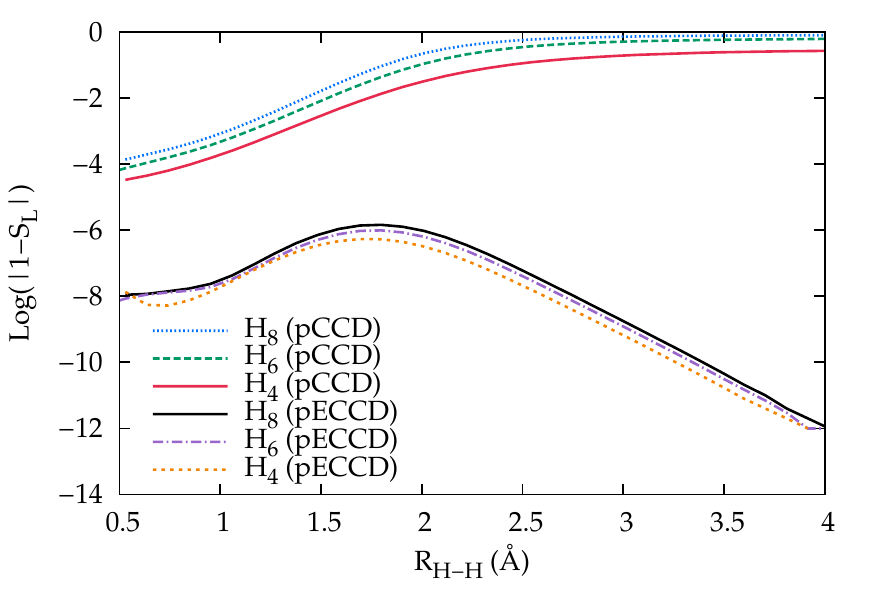}
\hfill
\includegraphics[width=0.48\textwidth]{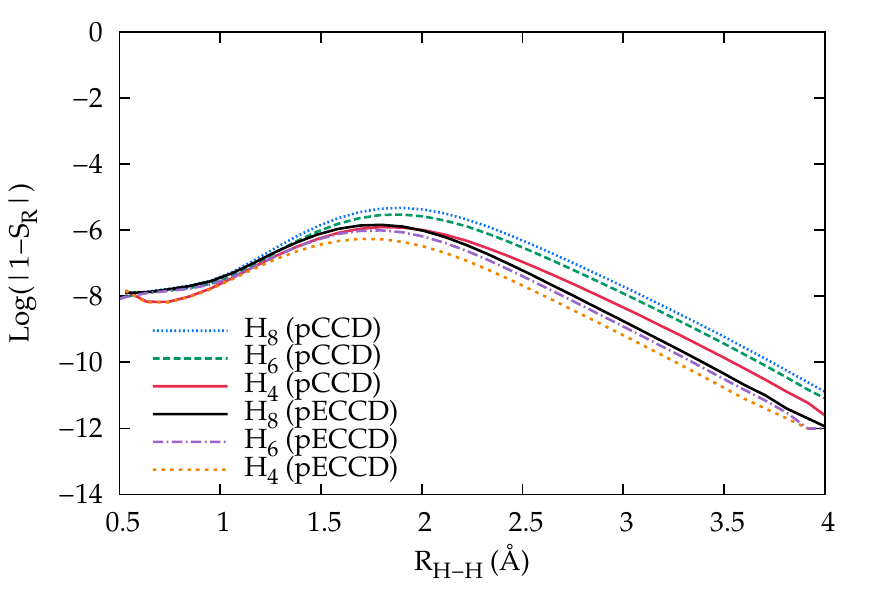}
\caption{Differences between pCCD, pECCD, and DOCI in equally spaced hydrogen chains.  Top left: Base 10 logarithm of the absolute value of $\Delta E$, measured in Hartrees and defined in Eqn. \ref{DefDE}.  Top right: Base 10 logarithm of the absolute value of $1-S$ as defined in Eqn. \ref{DefS}.  Bottom left: Base 10 logarithm of the absolute value of $1-S_L$ as defined in Eqn. \ref{DefSL}.  Bottom right: Base 10 logarithm of the absolute value of $1-S_R$ as defined in Eqn. \ref{DefSR}.  Kinks in the pCCD results for $\Delta E$ and $1-S$ are due to changes in sign.
\label{Fig:HChains}}
\end{figure}

We begin, then, with chains of equally spaced hydrogen atoms.  The strong correlations in these systems seem to be described reasonably well by DOCI.  Figure \ref{Fig:HChains} shows that while pCCD reproduces DOCI quite well energetically, pECCD does so even better.  More relevantly, while both pCCD and pECCD accurately describe the right-hand wave function (\textit{i.e.} we see that $|\mathrm{DOCI}\rangle \approx \exp(T) |0\rangle$), the left-hand wave function of pCCD is a fairly poor approximation to the DOCI state, while the left-hand wave function of pECCD is again very accurate.  This should not be too surprising, as the pCCD left-hand wave function $\langle 0| (1+Z) \, \exp(-T)$ consists only of the reference and doubly excited determinants, while the pECCD left-hand wave function includes all the same higher excitations that DOCI adds.  Indeed, that pECCD accurately reproduces the left-hand wave function of DOCI while pCCD sometimes does not seems to be a fairly general feature.  Note that while an accurate left-hand wave function is not needed for accurate energies, errors in the left-hand wave function may translate into errors for properties other than the energy.  In other words, while both pCCD and pECCD match the DOCI energy, we would expect only pECCD to match DOCI for arbitrary properties.  We should also perhaps emphasize the smallness of the various discrepancies; the pECCD energy differs from DOCI by less than 0.001 kcal/mol per electron, and the left- and right-hand states have overlaps with DOCI differing from 1 by about $10^{-8}$, implying that the coefficients of DOCI excited states in the pECCD ground state are less than $10^{-4}$.  For all practical purposes, pECCD reproduces DOCI exactly.

\begin{figure}[t]
\includegraphics[width=0.48\textwidth]{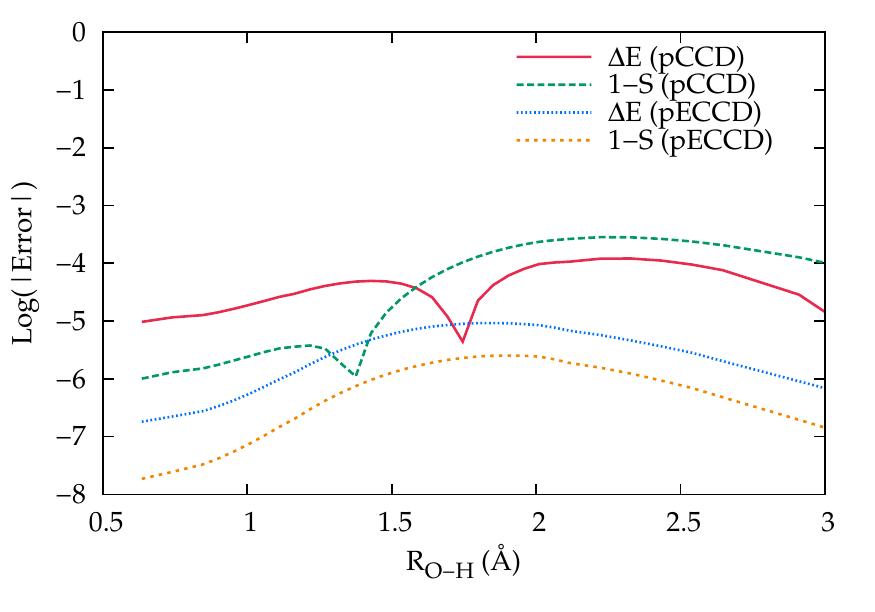}
\hfill
\includegraphics[width=0.48\textwidth]{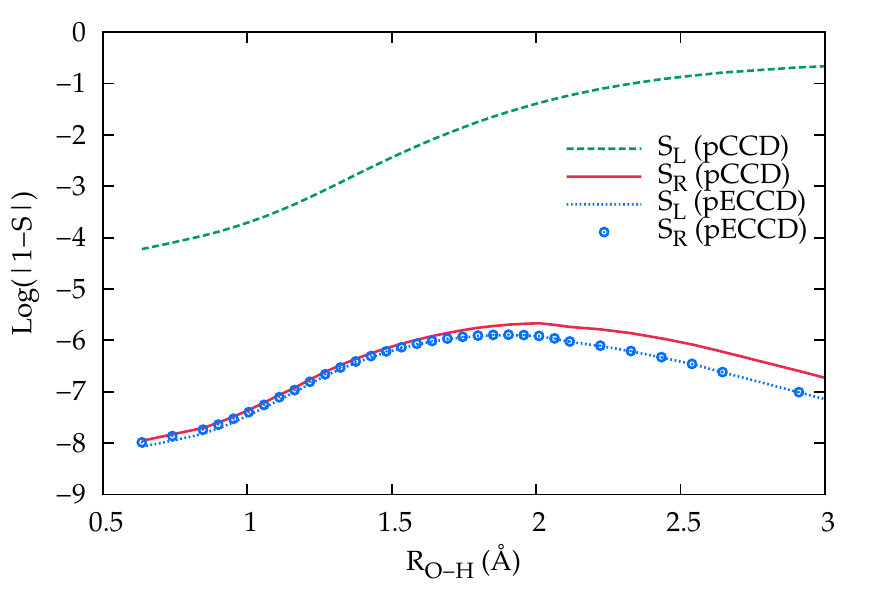}
\caption{Differences between pCCD, pECCD, and DOCI in the dissociation of H$_2$O.  Left panel: Base 10 logarithms of the absolute value of $\Delta E$ (measured in Hartrees and defined in Eqn. \ref{DefDE}) and of $1-S$ (defined in Eqn. \ref{DefS}).  Right panel: Base 10 logarithms of the absolute value of $1-S_L$ and $1-S_R$, defined in Eqn. \ref{DefSLR}.  Kinks in the pCCD results in the left panel are due to changes in sign.
\label{Fig:H2O}}
\end{figure}

\begin{figure}[t]
\includegraphics[width=0.48\textwidth]{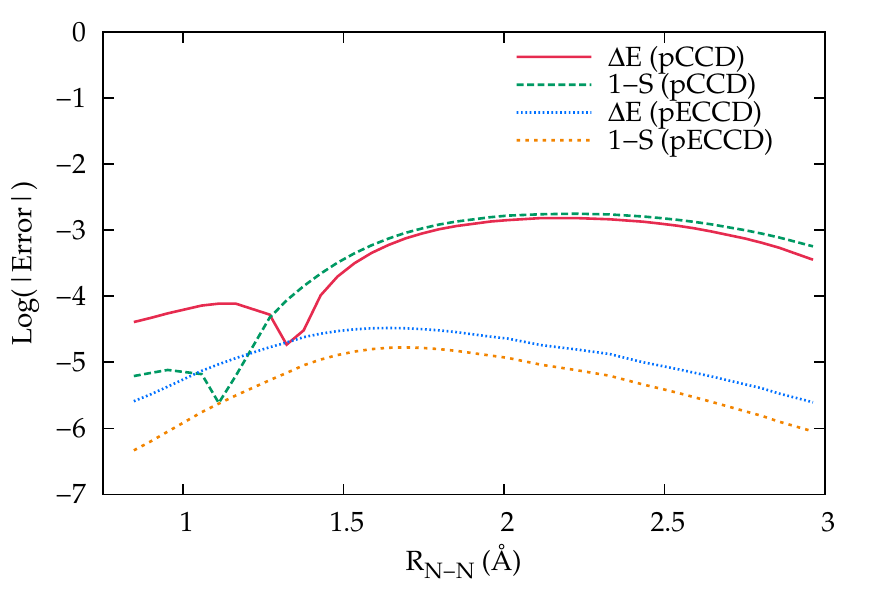}
\hfill
\includegraphics[width=0.48\textwidth]{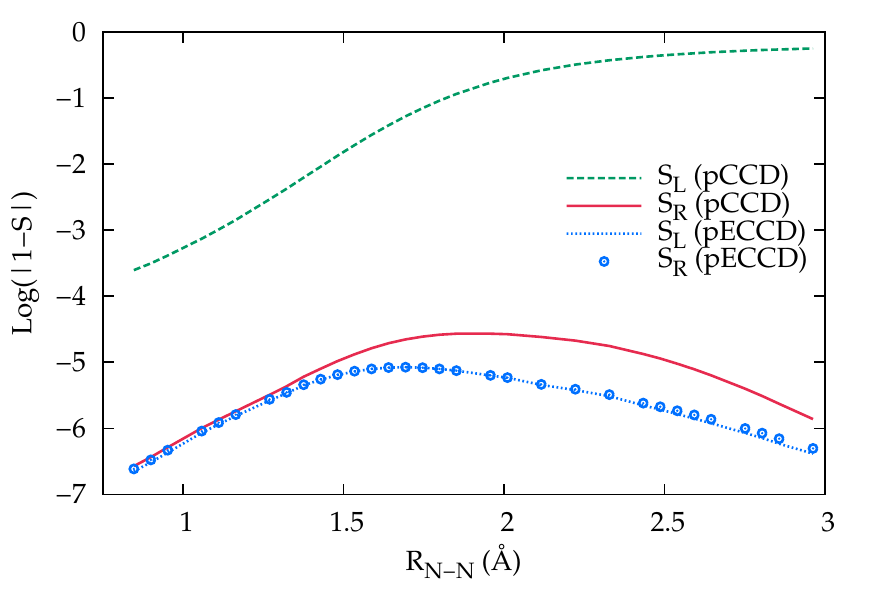}
\caption{Differences between pCCD, pECCD, and DOCI in the dissociation of N$_2$.  Left panel: Base 10 logarithms of the absolute value of $\Delta E$ (measured in Hartrees and defined in Eqn. \ref{DefDE}) and of $1-S$ (defined in Eqn. \ref{DefS}).  Right panel: Base 10 logarithms of the absolute value of $1-S_L$ and $1-S_R$, defined in Eqn. \ref{DefSLR}.  Kinks in the pCCD results in the left panel are due to changes in sign.
\label{Fig:N2}}
\end{figure}

\begin{figure}[t]
\includegraphics[width=0.48\textwidth]{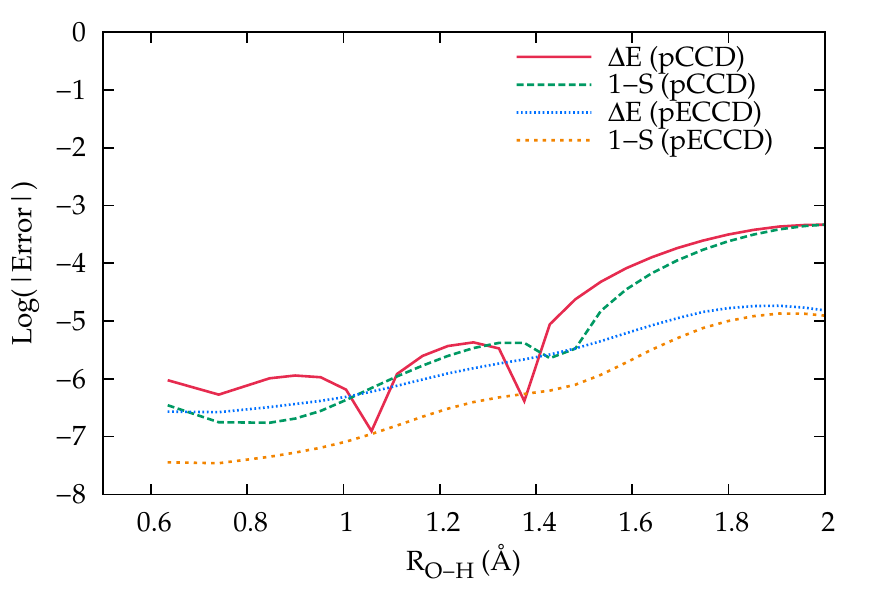}
\hfill
\includegraphics[width=0.48\textwidth]{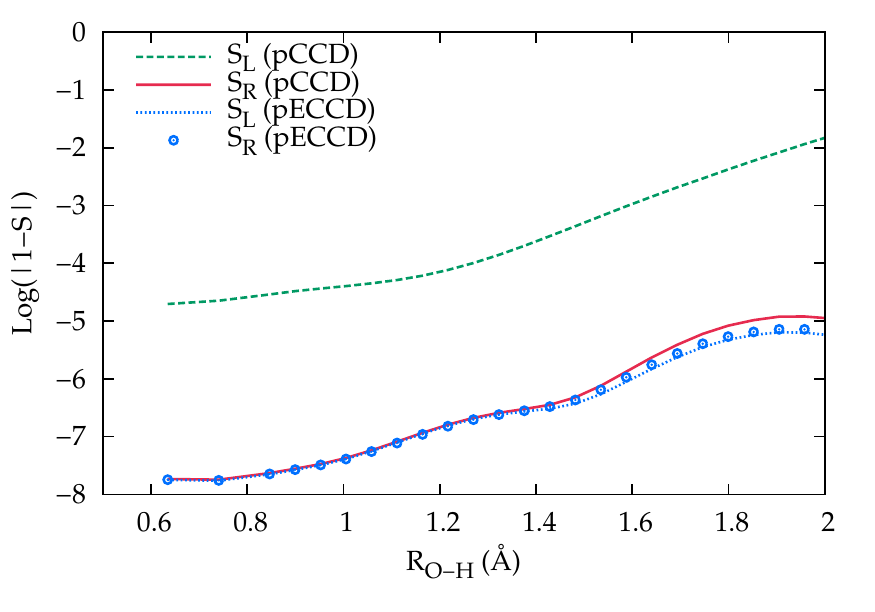}
\caption{Differences between pCCD, pECCD, and DOCI in the dissociation of H$_2$O, pairing canonical Hartree-Fock orbitals.  Left panel: Base 10 logarithms of the absolute value of $\Delta E$ (measured in Hartrees and defined in Eqn. \ref{DefDE}) and of $1-S$ (defined in Eqn. \ref{DefS}).  Right panel: Base 10 logarithms of the absolute value of $1-S_L$ and $1-S_R$, defined in Eqn. \ref{DefSLR}.  Kinks in the pCCD results in the left panel are due to changes in sign.
\label{Fig:H2OCan}}
\end{figure}

Similar results are seen in the symmetric double dissociation of H$_2$O (Fig. \ref{Fig:H2O}) and in the dissociation of N$_2$ (Fig. \ref{Fig:N2}).  While the pCCD energy and right-hand wave function are close to those of DOCI, pECCD is closer yet; meanwhile, the pECCD left-hand wave function may be a much better approximation to DOCI than is the pCCD left-hand wave function.  This is true not just in the basis of energetically optimized orbitals, but is also true with canonical Hartree-Fock orbitals (see, \textit{e.g.}, Fig. \ref{Fig:H2OCan}).

\begin{figure}[t]
\includegraphics[width=0.48\textwidth]{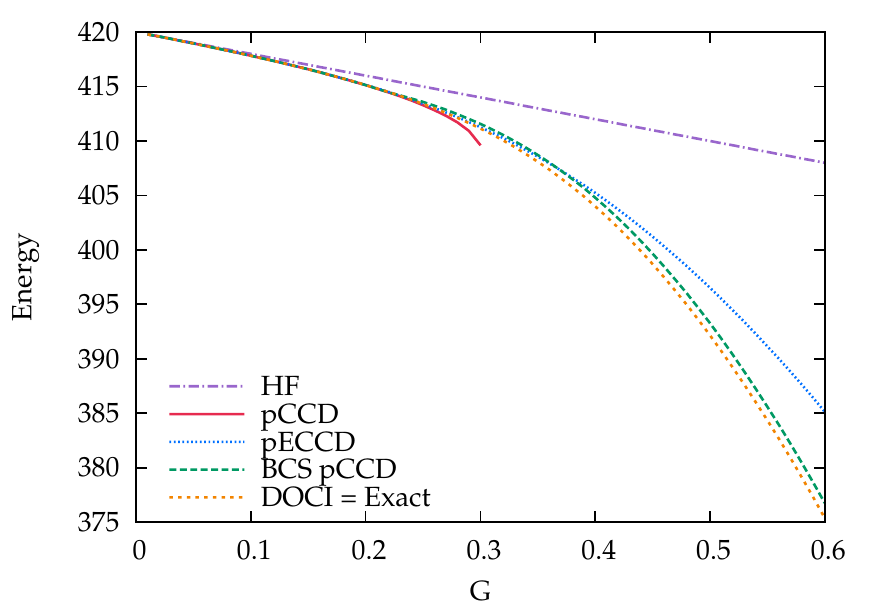}
\hfill
\includegraphics[width=0.48\textwidth]{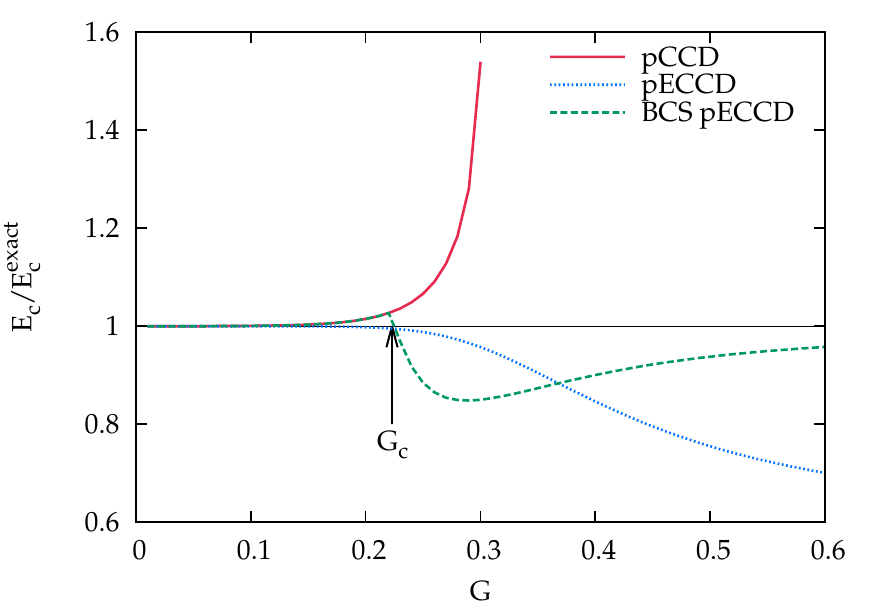}
\caption{Energies in the 40-site pairing Hamiltonian at half-filling.  Left panel: total energies.  Right panel: fraction of the DOCI correlation energy recovered by pCCD, pECCD, and by pCCD based on a number-broken BCS reference.  Near the value marked $G_c$, number symmetry is spontaneously broken at the mean-field level, as can be seen from the BCS-based pCCD results.
\label{Fig:PairingHam}}
\end{figure}

We have mentioned that the close coincidence between pCCD and DOCI breaks down for the attractive pairing Hamiltonian of Eqn. \ref{Eqn:PairingHam} which in the language of the SU(2) generators discussed earlier is just
\begin{equation}
H = \sum \epsilon_p \, N_p - G \, \sum_{pq} P_p^\dagger \, P_q.
\end{equation}
Because this Hamiltonian contains only the SU(2) pairing or pseudospin generators, it is solved exactly by DOCI.  Conveniently, however, a more compact exact solution was found by Richardson,\cite{Richardson1,Richardson4} which permits the generation of exact energies for systems far too large to be practicably solved by DOCI.  In Fig. \ref{Fig:PairingHam} we consider a 40-site pairing Hamiltonian with equally-spaced levels ($\epsilon_p = p$) at half-filling.  One can see that near $G \approx 0.2$, pCCD begins to deviate significantly from DOCI, and for $G \gtrsim 0.3$ we can find no real solution to the pCCD equations (solutions with complex $T$ amplitudes and complex correlation energies exist, but are of limited physical interest).  While pECCD also begins to break down somewhat, it provides a much more accurate desciption of the correlations in the pairing Hamiltonian for large $G$.  Note that for the half-filled forty-site Hamiltonian under consideration, a broken number symmetry mean field appears at $G_c \approx 0.22$, not coincidentally close to the value at which pCCD begins to break down.  While pECCD is not a panacea, it provides results much superior to pCCD for the pairing Hamiltonian and competitive with pCCD based on the broken-symmetry mean-field\cite{Henderson2014} for $G$ not too large.

\begin{figure}[t]
\includegraphics[width=0.48\textwidth]{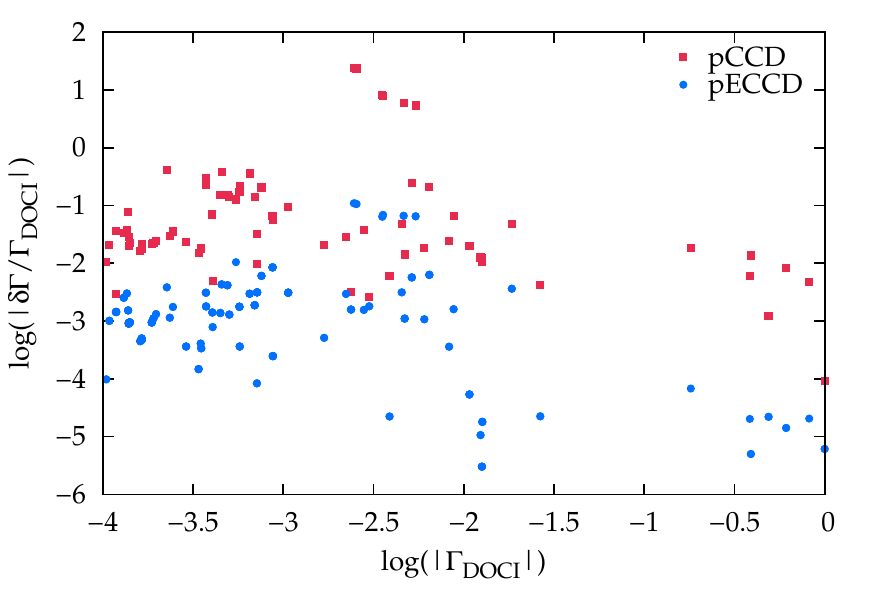}
\caption{Base 10 logarithms of the fractional differences between pCCD, pECCD, and DOCI two-particle density matrices.  Here, $\delta\Gamma$ means the difference between entries of the pCCD or pECCD density matrix and the DOCI density matrix.  We have omitted elements of the DOCI density matrix smaller in magnitude than $10^{-4}$.
\label{Fig:N2DMat}}
\end{figure}

As discussed earlier, because the pECCD left-hand wave function has much better overlap with the DOCI wave function that does the left-hand wave function of pCCD, one would expect pECCD to yield density matrices closer to the DOCI density matrices than does pCCD.  This is indeed the case.  Figure \ref{Fig:N2DMat} shows fractional errors in the entries of the two-particle density matrix contributions $\langle P_p^\dagger \, P_q \rangle$ for N$_2$ in the recoupling regime ($R_{N-N} \approx 2.2$ \AA).  Explicitly, we are plotting the fractional error $1 - \Gamma/\Gamma_\mathrm{DOCI}$ in the matrix elements of the two-particle density matrix for all elements $\Gamma_\mathrm{DOCI}$ larger than $10^{-4}$, since very small elements of $\Gamma_\mathrm{DOCI}$ are unlikely to have significant contributions to expectation values.  We see that typically, the pCCD values differ from those of DOCI by $\sim 1\%$, while the errors in the pECCD density matrix elements are an order of magnitude or more smaller.  Note that we have Hermitized the pCCD and pECCD density matrices to simplify the comparison; this has a much larger effect on pCCD than on pECCD because the pECCD density matrix is more nearly Hermitian.  For example, the Frobenius norms $\sqrt{\mathrm{Tr}(\Gamma \, \Gamma^\mathsf{T})}$ of the antisymmetric parts of two-particle density matrices discussed in Fig. \ref{Fig:N2DMat} are $5.15 \times 10^{-2}$ for pCCD and $7.68 \times 10^{-5}$ for pECCD.

\section{Conclusions
\label{Sec:Conclusions}}
In many cases, a configuration interaction restricted to paired excitations but not restricted by excitation level can provide an accurate accounting for static correlation effects, particularly once the orbitals used to define the pairing and the reference determinant are optimized.  The idea that this doubly occupied configuration interaction could describe many forms of static correlation is not a new one; indeed, DOCI itself was introduced over forty years ago.\cite{Allen1962,Smith1965,Weinhold1967,Veillard1967}  But while pair-excited configuration interaction attracted a great deal of early interest, the method was basically abandoned due to its exponential scaling with system size, which renders DOCI unsuitable for practical calculations.

With pair coupled cluster doubles and pair extended coupled cluster doubles, we now possess two models in the coupled cluster family which generally provide results almost indistinguishable from those of DOCI but with mean-field computational cost.  While the pCCD energy and right-hand wave function typically reproduce DOCI very well, the left-hand wave function, being a linear combination of the reference and doubly-excited determinants, is not always of particularly good quality, and the coincidence between pCCD and DOCI breaks down for the attractive pairing interaction (though note that the coefficient of the pairing-type interaction in the physical electronic Hamiltonian is positive, and can only become negative upon renormalization with some Hamiltonian transformation).  Both these difficulties can be ameliorated by using pair extended coupled cluster doubles, which seems to offer much more accurate results for attractive interactions and a much superior left-hand wave function (and therefore superior density matrices and expectation values).  The computational scaling of pECCD is the same as pCCD, namely $\mathcal{O}(N^3)$, though the pECCD energy and amplitude equations are significantly more complicated and the method is accordingly rather more expensive than is pCCD itself.  Nonetheless, the mean-field computational scaling permits routine pECCD calculations on systems with hundreds of basis functions, for which we can reliably anticipate getting results of essentially DOCI quality for both the energy and for other observables.  We hope, then, that pECCD will be a valuable tool for the description of strongly correlated systems, particularly when pairing interactions become attractive and $G$ is not too large, and that it will form an excellent starting point from which to add correlations which break pairs.

There are, of course, important drawbacks of pECCD as well.  The most significant is that pECCD does not include dynamic correlation.  One can attempt to fix this with the addition of a simple density functional correlation energy,\cite{Garza2015} or by relaxing the restriction to seniority zero, whether by freezing amplitudes\cite{Stein2014,Henderson2014b} or, potentially, by including a perturbative account for higher seniority sectors.  Moreover, the restriction to seniority zero is sometimes too severe; some systems simply require higher seniority sectors for the description of strong correlation effects, as is seen, for example, in the dissociation of N$_2$.\cite{Bytautas2011}  We may hope to treat these problems by lifting the restriction that the pairing scheme pairs the two spinorbitals corresponding to the same spatial orbital.  This might even open the door to the use of broken spin symmetry in seniority zero methods for strongly correlated systems.

\section{Acknowledgments}
This work was supported by the National Science Foundation (CHE-1462434).  GES is a Welch Foundation chair (C-0036).  We thank Jorge Dukelsky for helpful discussions.

\appendix
\section{Useful Derivations}
Here, we sketch a few derivations which the reader may find helpful but which are not necessary for understanding the thrust of the manuscript.

\subsection{Derivation of $H_0^{\delta\Omega=0}$}
Recall that $H_0^{\delta\Omega=0}$ is the portion of the Hamiltonian which preserves seniorities of individual spatial orbitals.  Here, we wish to derive the expression quoted in Eqn. \ref{DefH0}.

Let us start, then, with the one-electron part of the Hamiltonian.  In order to preserve orbital seniority, when we remove an electron from orbital $p$ we must return it to orbital $p$.  Thus, we have
\begin{subequations}
\begin{align}
\sum_{pq} \sum_\sigma \langle p | h | q \rangle \, a_{p_\sigma}^\dagger \, a_{q_\sigma}
 \to &\sum_p \langle p | h | p \rangle \sum_\sigma  a_{p_\sigma}^\dagger \, a_{p_\sigma}
\\
   = &\sum_p \langle p | h | p \rangle \, \sum_\sigma n_{p_\sigma}
\\
   = & \sum_p \langle p | h | p \rangle \, N_p
\end{align}
\end{subequations}
where the spinorbital number operators are
\begin{equation}
n_{p_\sigma} = a_{p_\sigma}^\dagger \, a_{p_\sigma}
\end{equation}
and their sum is the spatial orbital number operator $N_p$ given in Eqn. \ref{DefN}.

The two-electron part of the Hamiltonian is slightly more complicated.  We could remove two electrons from orbital $p$ and place them in orbital $q$, or we could remove one electron from orbital $p$ and another from orbital $q \neq p$, in which case we must place electrons back in orbitals $p$ and $q$; in the latter case, we must include the possibility of an exchange where the first electron is removed from $p$ but placed in $q$.  All told, we have
\begin{subequations}
\begin{align}
\frac{1}{2} \, \sum_{pqrs} \, \sum_{\sigma\eta} \langle pq| v | rs \rangle \, a_{p_\sigma}^\dagger \, a_{q_\eta}^\dagger \, a_{s_\eta} \, a_{r_\sigma}
 \to &\frac{1}{2} \, \sum_{pq} \, \sum_{\sigma\eta} \langle pp| v | qq \rangle \, a_{p_\sigma}^\dagger \, a_{p_\eta}^\dagger \, a_{q_\eta} \, a_{q_\sigma}
\\
   + &\frac{1}{2} \, \sum_{p \neq q} \, \sum_{\sigma\eta} \langle pq| v | pq \rangle \, a_{p_\sigma}^\dagger \, a_{q_\eta}^\dagger \, a_{q_\eta} \, a_{p_\sigma}
\nonumber
\\
   + &\frac{1}{2} \, \sum_{p \neq q} \, \sum_{\sigma\eta} \langle pq| v | qp \rangle \, a_{p_\sigma}^\dagger \, a_{q_\eta}^\dagger \, a_{p_\eta} \, a_{q_\sigma}
\nonumber
\\
  =& \frac{1}{2} \, \sum_{pq} \langle pp| v | qq \rangle \, \sum_{\sigma}  a_{p_\sigma}^\dagger \, a_{p_{\bar{\sigma}}}^\dagger \, a_{q_{\bar{\sigma}}} \, a_{q_\sigma}
\\
  +& \frac{1}{2} \, \sum_{p \neq q} \langle pq| v | pq \rangle \, \sum_{\sigma\eta} n_{p_\sigma} \, n_{q_\eta}
\nonumber
\\
  -& \frac{1}{2} \, \sum_{p \neq q} \langle pq| v | qp \rangle \, \sum_{\sigma\eta} a_{p_\sigma}^\dagger \, a_{p_\eta} \, a_{q_\eta}^\dagger \, a_{q_\sigma}
\nonumber
\\
  =& \sum_{pq} \langle pp| v | qq \rangle \, P_p^\dagger \, P_q
\\
  +& \frac{1}{2} \, \sum_{p \neq q} \langle pq| v | pq \rangle \, N_p \, N_q
\nonumber
\\
  -& \frac{1}{2} \, \sum_{p \neq q} \langle pq| v | qp \rangle \,  \left(\sum_\sigma n_{p_\sigma} \, n_{q_\sigma} + S^+_p \, S^-_q + S^-_p \, S^+_q\right)
\nonumber
\end{align}
\end{subequations}
where the spin index $\bar{\sigma}$ is the opposite of the index $\sigma$ and where we have made use of the pair creation operator $P_p^\dagger$ given in Eqn. \ref{DefP} and the adjoint operator $P_q$ and have introduced the spin raising and lowering operators
\begin{subequations}
\begin{align}
S^+_p &= a_{p_\uparrow}^\dagger \, a_{p_\downarrow}
\\
S^-_p &= a_{p_\downarrow}^\dagger \, a_{p_\uparrow}.
\end{align}
\end{subequations}

To simplify further we note that
\begin{subequations}
\begin{align}
n_{p_\uparrow} &= \frac{1}{2} \, N_p + S^z_p,
\\
n_{p_\downarrow} &= \frac{1}{2} \, N_p - S^z_p
\end{align}
\end{subequations}
where
\begin{equation}
S^z_p = \frac{1}{2} \, \left(a_{p_\uparrow}^\dagger \, a_{p_\uparrow} - a_{p_\downarrow}^\dagger \, a_{p_\downarrow}\right) = \frac{1}{2} \, \sum_{\mu\nu} a_{p_\mu}^\dagger \, \sigma^z_{\mu\nu} \, a_{p_\nu}
\end{equation}
and $\bm{\sigma}^z$ is a Pauli matrix.  This means that
\begin{equation}
\sum_\sigma n_{p_\sigma} \, n_{q_\sigma} = \frac{1}{2} \, N_p \, N_q + 2 \, S_p^z \, S_p^q
\end{equation}
whence
\begin{subequations}
\begin{align}
\frac{1}{2} \, \sum_{pqrs} \, \sum_{\sigma\eta} \langle pq| v | rs \rangle \, a_{p_\sigma}^\dagger \, a_{q_\eta}^\dagger \, a_{s_\eta} \, a_{r_\sigma}
 &\to \sum_{pq} \langle pp| v | qq \rangle \, P_p^\dagger \, P_q
  + \frac{1}{4} \, \sum_{p \neq q} \left(2 \, \langle pq| v | pq \rangle - \langle pq| v | qp \rangle\right) \, N_p \, N_q
\\
 &- \frac{1}{2} \, \sum_{p \neq q} \langle pq| v | qp \rangle \,  \left(2 \, S_p^z \, S_p^q + S^+_p \, S^-_q + S^-_p \, S^+_q\right)
\nonumber
\\
 &= \sum_{pq} v_{pq} \, P_p^\dagger \, P_q
  + \frac{1}{4} \, \sum_{p \neq q} w_{pq} \, N_p \, N_q
  + \sum_{p \neq q} K_{pq} \, \vec{S}_p \cdot \vec{S}_q,
\end{align}
\end{subequations}
in terms of the spin vector operators defined in Eqn. \ref{DefSpin} so that overall, we may make the replacement
\begin{equation}
H \to \sum_p h_p \, N_p + \frac{1}{4} \, \sum_{p \neq q} w_{pq} \, N_p \, N_q + \sum_{pq} v_{pq} \, P_p^\dagger \, P_q + \sum_{p \neq q} K_{pq} \, \vec{S}_p \cdot \vec{S}_q,
\end{equation}
as desired.

\subsection{Density Matrices}
In order to compute the pECCD energy, we need to evaluate density matrix elements.  In other words, we need to take the expectation value
\begin{equation}
\langle H_0^{\delta\Omega=0} \rangle_\mathrm{pECCD} = \langle 0| \mathrm{e}^Z \, \mathrm{e}^{-T} \, H_0^{\delta \Omega=0} \, \mathrm{e}^T |0\rangle,
\end{equation}
where we recall that $|0\rangle$ is the single-determinant reference.  As noted in Ref. \onlinecite{Henderson2014b}, the density matrices of seniority zero methods are sparse, and in fact the only non-zero entries of the full one- and two-particle density matrices can be constructed from $\langle N_p \rangle_\mathrm{pECCD}$, $\langle N_p \, N_q \rangle_\mathrm{pECCD}$, and $\langle P_p^\dagger \, P_q \rangle_\mathrm{pECCD}$.

Computing the density matrix elements is simplified by constructing similarity transformations of the number operators $N_p$ and the pair creation and annihilation operators $P_p^\dagger$ and $P_q$.  We find
\begin{subequations}
\begin{align}
\mathrm{e}^{-T} \, N_a \, \mathrm{e}^T
 &= N_a + 2 \, \sum_i t_{ia} \, P_a^\dagger \, P_i,
\\
\mathrm{e}^{-T} \, N_i \, \mathrm{e}^T
 &= N_i - 2 \, \sum_a t_{ia} \, P_a^\dagger \, P_i,
\\
\mathrm{e}^{-T} \, P_a^\dagger \, \mathrm{e}^T
 &= P_a^\dagger,
\\
\mathrm{e}^{-T} \, P_i^\dagger \, \mathrm{e}^T
 &= P_i^\dagger + \sum_{b} t_{ib} \, P_b^\dagger \, (N_i - 1) - \sum_{a \neq b} t_{ia} \, t_{ib} \, P_a^\dagger \, P_b^\dagger \, P_i,
\\
\mathrm{e}^{-T} \, P_a \, \mathrm{e}^T
 &= P_a + \sum_{i} t_{ia} \, (1 - N_a) \, P_i - \sum_{i \neq j} t_{ia} \, t_{ja} \, P_a^\dagger \, P_i \, P_j,
\\
\mathrm{e}^{-T} \, P_i \, \mathrm{e}^T
 &= P_i,
\end{align}
\label{DefOBar}
\end{subequations}
where we have taken advantage of the commutator expansion
\begin{equation}
\mathrm{e}^{-T} \, \mathcal{O} \, \mathrm{e}^T = \mathcal{O} + [\mathcal{O},T] + \frac{1}{2} \, [[\mathcal{O},T],T] + \frac{1}{3!} \, [[[\mathcal{O},T],T],T] + \ldots
\end{equation}
and have used the SU(2) commutation relationships given in Eqn. \ref{DefSU2} and transcribed here for convenience:
\begin{subequations}
\begin{align}
[P_p, P_q^\dagger] &= \delta_{pq} \, \left(1 - N_p\right),
\\
[N_p, P_q] &= -2 \, \delta_{pq} \, P_q.
\end{align}
\end{subequations}
We have added a few summation index restrictions which are unnecessary (but permitted) in view of the nilpotency of the pair creation and annihilation operators, simply because they may help clarify the origins of the summation restrictions and factors such as $\bar{\delta}_{ab}$ and $\bar{\delta}_{ij}$ in the final density matrices of Eqn. \ref{DefDMats}.

\bibliography{pECCD}

\end{document}